\documentclass[12pt]{iopart}
\usepackage{graphicx}
\usepackage{epstopdf}
\usepackage{multirow}
\usepackage{subfig}
\usepackage{cite}
\usepackage{float}

\begin{document}

\title[]{A Novel Quantum Algorithm for Ant Colony Optimization}

\author{Qian Qiu$^{*b}$, Mohan Wu$^{*a}$, Qichun Sun$^b$, Xiaogang Li$^b$, Hua Xu$^{a\dagger}$}
\address{$^a$College of Artificial Intelligence, Tianjin University of Science and Technology, Tianjin 300457, China}
\address{$^b$ Yiwei Quantum Technology Co., Ltd, Hefei 230088, China}
\ead{hua.xu@ywquantum.com (H. Xu)}
\vspace{10pt}
\begin{indented}
\item[]February 2024
\end{indented}

\footnote{* The first and second author contributes equally to this work.}
\begin{abstract}
Quantum ant colony optimization (QACO) has drew much attention since it combines the advantages of quantum computing and ant colony optimization (ACO) algorithms and overcomes some limitations of the traditional ACO algorithm. However, due to the hardware resource limitations of currently available quantum computers, such as the limited number of qubits, lack of high-fidelity gating operation, and low noisy tolerance, the practical application of the QACO is quite challenging.
In this paper, we introduce a hybrid quantum-classical algorithm by combining the clustering algorithm with QACO algorithm, so that this extended QACO can handle large-scale optimization problems, which makes the practical application of QACO based on available quantum computation resource possible.
To verify the effectiveness and performance of the algorithm, we tested the developed QACO algorithm with the Travelling Salesman Problem (TSP) as benchmarks. The developed QACO algorithm shows better performance under multiple data set. In addition, the developed QACO algorithm also manifests the robustness to noise of calculation process, which is typically a major barrier for practical application of quantum computers.  Our work shows that the combination of the clustering algorithm with QACO has effectively extended the application scenario of QACO in current NISQ era of quantum computing.
\end{abstract}

%
\vspace{2pc}
\noindent{\it Keywords}: Quantum Computing, Hybrid quantum algorithm, Quantum ant colony optimization, Travelling Salesman Problem
%
%
%
%

\section{Introduction}
Since the Italic scholars, A. Colorni and M. Dorigo, proposed the Ant Colony Optimization (ACO) in the 1990s, the heuristic algorithm is suitable for solving optimization issues, emulating the ant colony foraging process in the wild environment\cite{Dorigo}. During the foraging process, each ant in the colony tries to find the shortest path to the food source based on the indirect communication among individuals, it is called pheromone in the biological terminology\cite{Birattari1}. In details, every foraging ant tries to find the shortest path way to the food site, leaving the bio-chemical substance, pheromone, on the ground during the whole food searching process. This swarm inspired meta-heuristic algorithm is usually utilized for solving NP-hard combinational problems, such as Travelling Salesman Problem (TSP)\cite{Marchese,Elloumi,Chitty}, Quadratic Assignment Problem\cite{Demiral,Amirhosein,Alfredo}, Vehicle Routing Problem\cite{Doerner,Yanfang,Patrick,Rizzoli} and many other issues.


In response to the demand for computing resources to solve NP-hard problems, Richard Feynman proposed a new computing architecture based on quantum systems, known as quantum computing, in 1982. In 1985, David Deutsch formulated the concept of a quantum algorithm, specifically the quantum parallelism and quantum interference that underpin quantum computation\cite{Deutsch}. Later in 1994, Peter Shor introduced his famous quantum algorithm for factoring large numbers, which demonstrated the potential of quantum computers to break classical cryptographic systems\cite{Shor}. Quantum computing capitalizes on the phenomenon of superposition, allowing quantum bits or qubits to exist in multiple states simultaneously. This vastly expands the potential for complex computations, including neural networks, genetic algorithms, and ant colony optimization\cite{Schmidhuber,Prieto,McCall,Khaled,Birattari}.

The advantages of quantum computation attract many researchers trying to modify ant colony optimization with quantum calculation. The often applied methods can be categorized into two ways, which are quantum algorithms and quantum inspired algorithm\cite{QNiu}.
In 2012, P. Li and H. Wang introduced a novel Quantum-Inspired Ant Colony Optimization (QIACO) utilizing the Bloch spherical search approach in quantum computation\cite{LiP}.
Meanwhile, some hybrid ACO algorithm have also been proposed recently, usually involving a combination of quantum and classical components in one algorithm. In 2022, M. Garcia de Andoin and J. Echanobe proposed an hybrid quantum ACO algorithm that was using quantum computing for tackling pheromone updating and exploration parameter, then passing the processing result to a classical computer for the rest parties of this algorithm\cite{Echanobe}.
However, due to the limitation of quantity and quality of the available qubits of real quantum computer, the usefulness and practical application of the algorithm has not been tested and verified by large problems\cite{Echanobe}.

In this article, we propose a scalable hybrid quantum classical algorithm that combines the ant colony optimization algorithm with the K-means clustering method. This approach can partially overcome the hardware limitations of currently available real quantum computers and is capable of solving problems of a substantial scale, with potential for practical application.
The following parts of this article are organized in such a way. In the section 2, we review the recent literature concerning quantum ant colony optimization (QACO). In the section 3, we introduce the principle of classical and quantum version of ant colony optimization. In the section 4, we cover the explicit steps of our proposed QACO algorithm. In the section 5, we analyze the results of the QACO with K-means clustering method. In the final section, we summarize the conclusions and illustrate the potential research directions in the relevant field.

\section{Literature Review}
ACO is a active research field, with a wide range of applications such as vehicle routing, resource allocation, data clustering, image processing, and financial portfolio management, continuously attracting significant attention. For getting more insights into the ACO, we will review the literature concerning the QACO and QIACO. In the development of ACO, Neumann presented rigorous theoretical research in 2006, introducing an algorithm with only one ant\cite{Neumann}. The investigation of Neumann constructed some criterion that are widely used nowadays that updating pheromone when the solution of current iteration is better than the present best solution and introduced the pheromone regulation rule by the evaporation coefficient which indicated the impact degree between the the solution of current iteration and previous best solution, which would degrade the performance of ACO beneath some threshold of the evaporation factor. Furthermore, some researchers attempted to take the advantages of quantum computation, like the superposition, entanglement and coherence, to optimize the ant colony algorithm. The QACO algorithm unlike classical ACO, it is used quantum computational manipulations, the quantum bit (qubit) and quantum gate, to exert operations within the ant colony optimization algorithm. Ling Wang et al first proposed a novel pheromone updating strategy that utilized quantum rotation gates to refresh the pheromone depending on the iterative result in 2007\cite{QNiu}. In their work, they focused on the prompting the convergence performance and improving the premature situation that the iteration results stranding around the local optima. In 2010, Panchi Li et al proposed a quantum ant colony optimization that codified the position of each ant with many qubits\cite{Panchi}. The algorithm collected the local best solution depending on pheromone trail on each path by utilizing quantum gate to update every ant's representative qubits. To induce path mutation and extend the capacity of the solution pool, this algorithm import the quantum non-gates operations. The innovative point of the algorithm was to add the fitness function of the present position of the iteration with updating the pheromone. Then, the best path would incorporate the greater fitness function such that accelerated the guaranteed convergence searching process. In 2022, M. Garcia de Andoin and J. Echanobe proposed a hybrid quantum ant colony optimization algorithm embracing the updating pheromone and exploration parameter codified with qubits\cite{Echanobe}, subrogating the rest parties to the classical computer. This modification aimed to make QACO suitable for near-term quantum computers. However, it is noteworthy that this algorithm still requires a considerable number of qubits, posing a challenge when applied to resolve meaningful issues.

\section{Original ACO and Quantum Computing Circuits}
\subsection{Ant Colony Optimization Algorithm}
The ACO algorithm is a meta-heuristic inspired by the foraging behavior of ants. Its core principle lies in emulating the ability of ant colonies to find the shortest paths between their nest and food sources. As individual ants explore their surroundings, they deposit pheromone trails. These trails function as communication channels, guiding other ants toward the most promising paths. The ACO algorithm replicates this natural behavior by utilizing a population of artificial ants to search for optimal solutions within a given problem space.

In ACO, the problem is represented as a graph, with nodes represent the problem's components (e.g., cities in the Traveling Salesman Problem), and edges represent connections or paths between the components. Each ant in the population generates a solution by iteratively moving from one component to another according to a set of rules. The key component in ACO is the pheromone matrix, which is an abstraction of the pheromone trails left by real ants. The matrix keeps track of the pheromone levels on each edge of the graph. Initially, the pheromone levels are initialized uniformly or randomly. During the process of solution searching, ants determine the next component to visit based on the probabilities of a combination of heuristic information and the pheromone levels of the available edges. The heuristic information guides the ants towards promising components depending on some problem-based knowledge. The pheromone levels, on the other hand, represent the collective knowledge accumulated by the ant colony. As ants traverse edges, they deposit pheromone according to the quality of their solutions. The pheromone evaporation mechanism ensures that pheromone levels gradually decay over time, preventing premature convergence to local optimal solutions. The global update rule is another essential aspect of ACO. After each iteration, the pheromone levels are globally updated to reflect the quality of the solutions searching. This updating rule allows the best ant or a combination of multiple ants to deposit additional pheromone on the edges of the best solution(s) found so far. By reinforcing the paths of the best solutions, the algorithm biases the exploration towards prospective regions of the searching space, thus gradually converging towards the better solutions.

The foregoing mentioned process of updating pheromone can be summarized as the ant decision rule, which is using a series of rules to guide an ant to choose the next feasible node stochastically. The $k$th individual ant at $t$ moment located on the component $r$ shift to the subsequent component $s$ with the rule as below.

\begin{equation}
s = \left \{
\begin{array}{ll}
  \mathop{argmax}\limits_{ru\in allowed_k(t)} \left [ \tau_{ru}(t)^{\alpha} \eta_{ru}^{\beta } \right ] &\mbox{when}\ q\le q_0 \\
  S &\mbox{otherwise}
\end{array}
\right .
\end{equation}

Where $\tau_{ru}(t)$ indicated the value of the corresponding pheromone at moment $t$, $\eta_{ru}$ demonstrated the explicit of the specific problem requirement, $\alpha$ illustrated the impact factor, $q$ is a number that is uniformly random selected in the $[0, 1]$, $q_0$ is a constant number, $allowed_k(t)$ denoted the feasible component set by the k-th ant, $S$ is a parameter that was selected from $allowed_k(t)$ based on the following equation.

\begin{equation}
P_{rs}^k(t)= \left \{
\begin{array}{ll}
\frac{\tau_{rs(t)}^{\alpha}\eta_{rs}^{\beta}}{\sum\limits_{ru\in allowed_k(t)}\tau_{rs}(t)^{\alpha}\eta_{ru}^{\beta}} &\mbox{if}\ s \in allowed_k(t)\\
0 &\mbox{othewise}
\end{array}
\right.
\end{equation}

The pheromone evaporation factor in the algorithm represent the phenomenon in the real ant colony that the magnitude of pheromone deposited on the trail will be decayed over time, which also prevent the searching process from being trapped in the local optima.

\subsection{Quantum Computing Circuits}
Quantum computing is the technology which is built up on the principle of quantum physics, enabling it possessing the better parallelism than the classical computing architecture that is suitable for resolving the problem requiring searching large solutions pool. In the quantum computing field, the smallest unit which corresponded to the bit in classical computing area is called a qubit that is defined as $\left[ \alpha, \beta \right ]^T$. The $\alpha$ and $\beta$ are two complex numbers satisfying $\alpha^2+\beta^2=1$, whose norm indicates the probability of staying in the specific state. For instance, $\left | \alpha \right |^2$ denotes the probability of staying in '$1$' state. A series of states could be represented as:

\begin{equation}
x_i =
\left[
\begin{array}{c|c|c|c}
\alpha_1 & \alpha_2 & \dots   & \alpha_m \\
\beta_1  & \beta_2  & \dots   & \beta_m
\end{array}
\right]
\end{equation}
where $| \alpha_i |^2 + |\beta_i|^2=1$, $i=1,2,3,\dots,m$.

For instance, the system Q embracing $3$ qubits could be represented as:

\begin{equation}
Q =
\left[
\begin{array}{c|c|c}
\frac{\sqrt{2}}{2} & \frac{\sqrt{2}}{2} & \frac{1}{2} \\
\frac{\sqrt{2}}{2} & -\frac{\sqrt{2}}{2} & \frac{\sqrt{3}}{2}
\end{array}
\right]
\end{equation}

Then, the probabilities of system Q could be described like:
\begin{center}
$\frac{1}{4} \left | 000  \right \rangle +  \frac{\sqrt{3}}{4} \left | 001  \right \rangle - \frac{1}{4} \left | 010  \right \rangle - \frac{\sqrt{3}}{4} \left | 011  \right \rangle + \frac{1}{4} \left | 100  \right \rangle + \frac{\sqrt{3}}{4} \left | 101  \right \rangle - \frac{1}{4} \left | 110  \right \rangle - \frac{\sqrt{3}}{4} \left | 111  \right \rangle$
\end{center}
Which indicates the probabilities of the system Q staying in states $\left |000\right \rangle$, $|001>$, $\left |010\right \rangle$, $\left |011 \right \rangle$, $\left |100\right \rangle$, $\left |101 \right \rangle$, $\left |110\right \rangle$ and $\left |111 \right \rangle$ is $1/16$, $3/16$, $1/16$, $3/16$, $1/16$, $3/16$, $1/16$, $3/16$, correspondingly. This feature also illustrates that the $3$ qubits incorporating the information of $8$ states, that is more efficient than using the classical bits to denote system states.

In quantum computing, it is usually used the quantum gate to apply operations on the qubit to get a desired results. In general, a quantum rotation manipulation, $R(\theta)$ is deployed to operate the individual qubit as below:
\begin{equation}
\left[\begin{array}{cc}\alpha_{id}^{'} \\ \beta_{id}^{'}\end{array}\right] = R(\theta_{id})\left[\begin{array}{cc}\alpha_{id} \\ \beta_{id}\end{array}\right] = \left[\begin{array}{cc}\cos(\theta_{id})-\sin(\theta_{id}) \\ \sin(\theta_{id})-\cos(\theta_{id})\end{array}\right]\cdot \left[\begin{array}{cc}\alpha_{id} \\ \beta_{id}\end{array}\right]
\end{equation}
Where $\theta_{id}$ is the rotation angle. Another common used quantum gate is called CNOT that essentially is controlled gate. Specifically, the CNOT gate requires at least two qubits. The one is the control qubit that determine the operation to exert on the objective when the preset condition is satisfied. The other is the target part qubit. The CNOT gate reverse the target qubit depending on the preset condition which is usually set as $\left|1\right \rangle$. The CNOT gate works on the target qubit like below.

\begin{equation}
\left |x \right \rangle \left |y \right \rangle \rightarrow \left | y \oplus x \right \rangle
\end{equation}

\begin{figure}[htbp]
\centering
\includegraphics[scale=0.8]{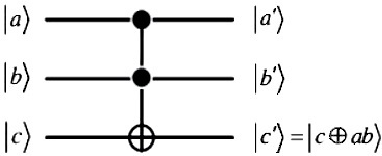}
\caption{The 2 controlled qubits CNOT gate (C$^2$NOT)}\label{C2NOT}
\end{figure}

The CNOT gate could also extend to a multiple controlled bits quantum gate if it is needed. The $2$ controlled qubits CNOT gate is shown in Fig. \ref{C2NOT}. This quantum gate could also be called as controlled-controlled NOT gate (C$^2$NOT).

\subsection{K-means Clustering}
The K-means clustering algorithm is a popular unsupervised machine learning technique used to partition data into distinct groups or clusters. Its simplicity, effectiveness, and interpretability have made it widely applied in various domains. The principle behind the K-means algorithm is to group data points based on their similarity, aiming to minimize the within-cluster sum of squares. The algorithm begins by randomly initializing K cluster centroids in the feature space, where K represents the desired number of clusters. Each data point is then assigned to the nearest centroid based on a distance metric, typically Euclidean distance. This initial assignment creates an initial clustering of the data.

Next, an iterative process starts to refine the clustering. The algorithm updates the centroids by computing the mean of all data points assigned to each cluster. This step recalculates the centroid positions, effectively shifting them towards the center of their respective clusters. Subsequently, the data points are reassigned to the closest centroids based on the updated positions. These centroid updates and reassignments continue iteratively until convergence, where the centroids no longer change significantly or a predetermined number of iterations is reached.

K-means clustering aims to minimize the within-cluster sum of squares, also known as the inertia or distortion. The inertia represents the sum of squared distances between each data point and its assigned centroid. By minimizing the inertia, the algorithm ensures that data points within each cluster are similar to each other, while distinct from data points in other clusters. The performance and effectiveness of the K-means algorithm heavily depend on the initial centroid positions. Different initializations can lead to different clustering results. To mitigate this issue, the algorithm is often run multiple times with different random initializations, and the clustering result with the lowest inertia is selected as the final solution.

K-means clustering has various applications in data analysis, pattern recognition, image segmentation, customer segmentation, and more. It provides a straightforward approach to uncovering underlying structures or patterns in data without requiring prior knowledge or labeled training samples. Despite its popularity, the K-means algorithm has certain limitations. One major challenge is determining the optimal number of clusters, K, which is often subjective and problem-dependent. Choosing an inappropriate value for K can lead to sub-optimal or misleading clustering results. Additionally, the algorithm is sensitive to outliers and can be influenced by the initial centroid positions, making it susceptible to converging to sub-optimal solutions.

To enhance the performance of K-means clustering, various extensions and variations have been proposed. These include initialization techniques like K-means plus to improve the initial centroid selection, and algorithms like K-means with mini-batch updates to handle large-scale datasets efficiently. Additionally, efforts have been made to address the limitations of K-means, such as incorporating robust distance metrics, handling categorical data, or incorporating constraints into the clustering process.
the K-means clustering algorithm is a widely used and versatile technique for unsupervised clustering. Its simplicity and interpretability make it accessible to practitioners in various domains. While it has its limitations, advancements and adaptations have been made to address these challenges, making K-means a valuable tool for data exploration, segmentation, and pattern recognition tasks.

\section{Quantum Ant Colony Optimization with K-means clustering method}
In recent years, it has gotten the magnificent advancements in quantum computing field. The application scope of quantum computing algorithm, however, is still confined due to the factors such as the scale of qubits and their fidelity. These limitations pose
significant challenges to the widespread adoption of quantum computing algorithms.

Generally, if the computational resources are limited, a common approach to resolve the large scale problem is to partition it into a smaller one that could be solved with available resources. This method could prompt the efficient utilization of resources and
tackle the large problem by breaking it into a relative small sub-problem. In this article, the main incentive that we import the K-means clustering method is to expand the application scope of the quantum ant colony optimization algorithm based on the current
existing quantum computing hardware. Decomposing a large problem into several relatively smaller problems is a natural idea, widely adopted in classical computing. However, when it comes to problems involving quantum algorithms, how to decompose them and what
the effectiveness is after decomposition are questions that are worth to explore.

The already existed QACO algorithm is confined mainly by the small scale of the qubits such that the QACO algorithm is often used as the demonstration that shows the virtue of the quantum computing. In our proposed modified QACO, the qubits are categorized into
two classes, the path generating qubits and ancilla qubit.

\begin{figure}[htbp]
\centering
\includegraphics[scale=0.5]{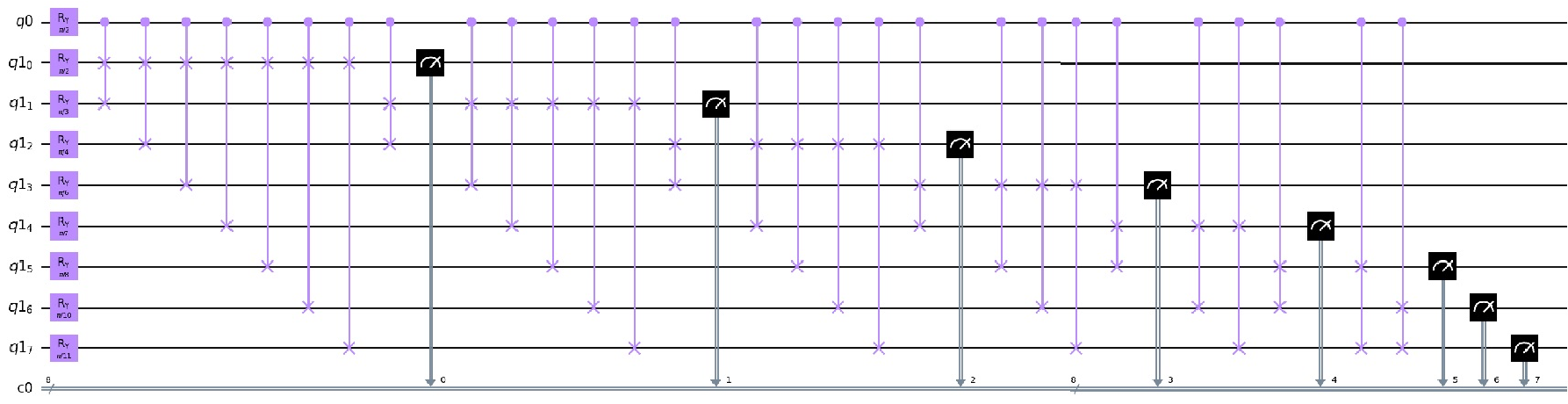}
\caption{The path search quantum circuit. In this diagram, $q0$ indicates the auxiliary qubit (ancilla) which make the solution to mutate in the certain condition controlled by Ry gate. Meanwhile, $q1_{0}$-$q1_{7}$ denote the possible path code with two qubits, e.g. $q1_{0}$ and $q1_{1}$ encodes the path $2$ as $10$ in binary.}\label{QCircuit}
\end{figure}

In our proposed QACO algorithm with K-means clustering, we utilize 9 qubits to process a 4 cities travelling salesman problem. The quantum circuit is shown in Fig. \ref{QCircuit}. The cities number are encode in binary using 2 qubits. Then the 8 qubits would
represent 4 cities.

\begin{table}[h]\centering
\caption{The total distances of the numerical tests by the ACO and QACO}\label{tab1}
\begin{tabular}{cccc}
\hline
                                            & \multicolumn{3}{c}{Parameters}           \\ \hline
Invariant times                             & $5$               &   $10$                   & $15$                   \\
AncillaryRotation Angle                     &   $\frac{\pi}{6}$        & $\frac{\pi}{4}$    & $\frac{\pi}{2}$  \\ \hline
\end{tabular}
\end{table}
\normalsize

In Fig. \ref{QCircuit}, $q0$ denote the auxiliary qubit (ancilla) which make the generated path to mutate when the global best path keeps invariant in many iterations. The Ry gate is used to control the mutation probability. The explicit is shown in
table \ref{tab1}. If the global best path kept invariant more than $15$ iterations, the ancilla mutation probability returns to zero.

As we know, the measurement of quantum computing is statistical. Since the outcome of the quantum circuit would generate the infeasible result which is not follow the rules of the TSP, the QACO must has a result examining module. If the current solution is
infeasible, the algorithm would generate a feasible path randomly in the first 10 iteration. After 10 iterations, the Hamming distance is imported into the algorithm. The probability of the infeasible solution is set to inversely proportional to the Hamming
distance from the existing solution pool. The probability is defined as below

\begin{equation}
p_i = \left(d_i\sum \limits_{j} \frac{1}{d_j}\right)^{-1}
\end{equation}

where, $d_i$ denotes the Hamming distance between the current infeasible solution and one of the best path in the solution pool, $j$ illustrates the number of the solution pool.

The main steps in the QACO with K-means clustering method can be described as below:\\
{\bf Step 1:} Utilizing K-means clustering to divide the larger problem into a smaller one that could be processed by the current quantum computing resource, where the value of the $k$ is relative to the applicable number of qubits.\\
{\bf Step 2:} Initializing the parameters of the QACO, setting the number of generations and the rotation angel of the pheromone to $\pi/2$, which means all solutions are generated with the same probabilities at the beginning of the algorithm.\\
{\bf Step 3:} Calculating the best solution in the iteration and comparing it with the global best solution. If the current best solution is better than the global best solution, then substitute with the current best solution.\\
{\bf Step 4:} Determining whether the current iteration meet the termination condition. If not, going to the next step.\\
{\bf Step 5:} Updating the pheromone value with the quantum rotation gate.\\
{\bf Step 6:} Determining whether the scale of the clustering group is suitable for the number of the qubit. If not, continuing to clustering and repeating the foregoing steps until the problem has be solved.

As the ACO, the updating pheromone strategy of QACO is also an important step. The updating pheromone table is given in table \ref{tab2}.

\begin{table}[h]\centering
\caption{The lookup table of the rotation angle of QACO}\label{tab2}
\begin{tabular}{cccc}
\hline
$x_i$ & $b_i$ & $f(x) > f(b)$ & $\Delta\theta_i$                   \\ \hline
0 & 0 & True                  & $-0.01\pi^{\ast}$ \\
0 & 0 & False                 & $0.04\pi $ \\
0 & 1 & True                  & $-0.05\pi^{\ast}$ \\
0 & 1 & False                 & $0.07\pi $ \\
1 & 0 & True                  & $0.05\pi^{\ast} $ \\
1 & 0 & False                 & $-0.07\pi$ \\
1 & 1 & True                  & $0.01\pi^{\ast} $ \\
1 & 1 & False                 & $-0.04\pi$ \\ \hline
\end{tabular}
\end{table}
\normalsize

where $x_i$ means the state of the i-th qubit on the current iteration, bi indicates the global best state of the i-th qubit so far. $f(x)$ and $f(b)$ demonstrates the value of the explicit fitness function based on the specific problem, the asterisk denotes the
cosine value of $\Delta\theta_i$ less than $-1$ then it must be multiplied with $-1$.
\section{The Numerical Results}
To test the performance and practicality of the QACO with K-means clustering method, it is employed Ulysses-16, bayg-29 and random generating data set to verify the features of the algorithm. The results of the tests are given below. Fig. \ref{ULYresults}.
provided the result of Ulysses-16. All the numerical tests are performed by Yiwei Quantum Computing Simulation Platform.

\begin{figure}[htbp]
\centering
\subfloat
{
\includegraphics[scale=0.5]{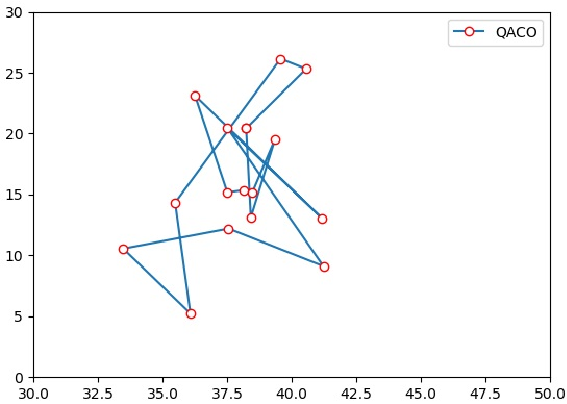}
}
\subfloat
{
\includegraphics[scale=0.5]{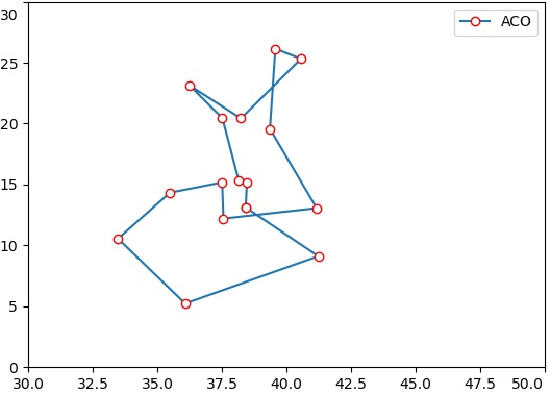}
}
\caption{The best path of the algorithm for solving Ulysses-16 (a) QACO (b) ACO}\label{ULYresults}
\end{figure}

The distance of this best path is 62.14 that is shorter than the standard distance of TSP which is 77.8372\cite{MeijiaoL}, shown in Fig. \ref{ULYresults}. (a). We also perform ACO algorithm on this data set with 6 ants, the pheromone and distance factor is 4 and 2, respectively. The total
number of the iterations is 1000. The best distance of the ACO is 69.67 that is also longer than the QACO with K-means clustering method. Meanwhile, to determine whether the K-means clustering making the progress of QACO, here, we calculate the Ulysses-16 by ACO
algorithm with K-means method, shown in Fig. \ref{ULYresults}. (b). The result of this method indicates the best distance is 66.34 that is longer than that of the QACO. It is indicated that the advancement of QACO is not induced by the clustering method.

\begin{figure}[htbp]
\centering
\includegraphics[scale=0.5]{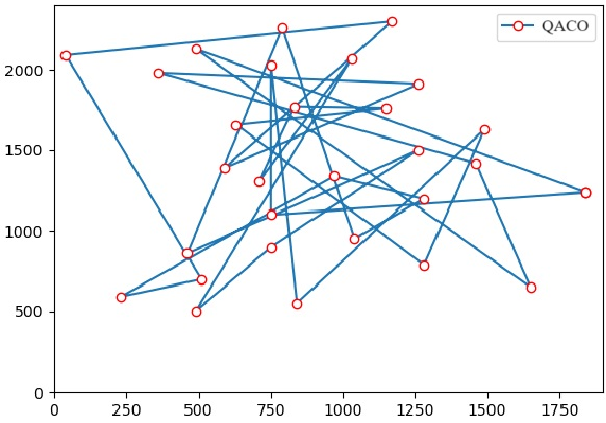}
\caption{The best path of the algorithm for solving bayg-29)}\label{BAYGresults}
\end{figure}

We also use the bayg-29 data set to verify the performance of the algorithm. The best path of the bayg-29 is shown in Fig. \ref{BAYGresults}.

The distance of this best path is 12545.95 that is longer than the standard distance of TSP which is 9407.35, which is acquired from Traveling Salesman Problem Library (TSPLIB) that is a library of sample instances for the TSP from
various sources. We also perform ACO algorithm on this data set with the same parameters set like before. The best distance of the ACO is 25089.08 that is also longer
than the QACO with K-means clustering method.

\begin{figure}[htbp]
\centering
\includegraphics[scale=0.5]{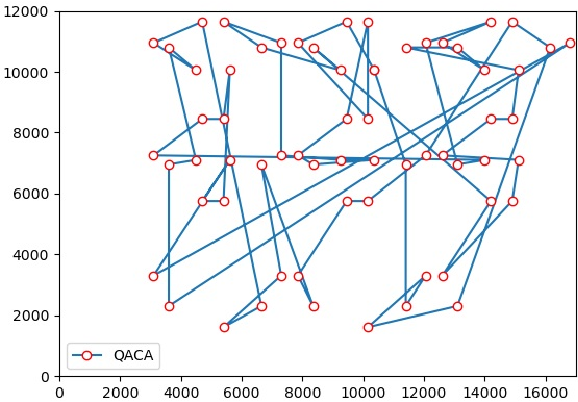}
\caption{The best path of the algorithm for solving random generating data set}\label{RANSETresults}
\end{figure}

Furthermore, the random generating data set with 64 points are shown in Fig. \ref{RANSETresults}. The best distance running by traditional ACO is 331187.30 and the result performing by the QACO is 148170.56. All numerical results are summarized in the following table, which
is shown in table \ref{tab3}.

\begin{table}[h]\centering
\caption{The total distances of the numerical tests by the ACO and QACO}\label{tab3}
\begin{tabular}{cccc}
\hline
     & Ulysses-16 & Bayg-29  & Random Data \\ \hline
ACO  & 69.67      & 25089.08 & 331187.30   \\
QACO & 62.14      & 12545.95 & 148170.56   \\ \hline
\end{tabular}
\end{table}
\normalsize

Since the current quantum computing hardware could not mitigate the effect of the noise, we also import 2 types of noise effect into the algorithm. The test results of the foregoing data set with noise effect are shown
in table \ref{NoiseTab}. The distances of the Bayg-29 that calculating with the two-qubit bit-flip noise and thermal relaxation noise by 0.1\%, 1\%, 2\%, 5\% and 10\% are close to the ideal simulation result.

\begin{table}[htbp]\centering
\caption{System parameters of some quantum computing prototypes}\label{tab_error}
\begin{tabular}{cccc}
\hline
Superconducting system  & Eagle\cite{IBMQ}       & Sycamore\cite{HeLH} & Zuchongzhi 2.1\cite{HeLH} \\ \hline
Number of qubits        & 127         & 53       & 66             \\
Single-qubit gate error & 0.15\%      & 0.16\%   & 0.16\%         \\
Two-qubit gate error    & 1.9\%\footnotemark[1]   & 0.62\%   & 0.60\%         \\ \hline\\
\end{tabular}
\end{table}
\footnotetext[1]{This is error per layered gate for a 100-qubit chain.}

In recent years, significant progress has been made in the development of quantum computing technology, with substantial advancements in areas such as the number of quantum bits and noise reduction, such as Eagle, Sycamore and Zuchongzhi 2.1 whose error parameters are shown in table \ref{tab_error} \cite{IBMQ, HeLH}.
The error testing range provided in the above table covers the two gate errors for current quantum computers, demonstrating that the algorithm can be applied on current quantum computing platforms.


\begin{table}[h]\centering
\caption{The results of Bayg-29 with noises.}\label{NoiseTab}
\begin{tabular}{cccccc}
\hline
                   & \multicolumn{5}{c}{Ratio of Noise} \\
Noise Model        & 0.1\%    & 1\%   & 2\%   & 5\%   & 10\%  \\ \hline
Bit-flip           & 14710.74 & 15137.13 & 13092.58 & 13941.60 & 12759.88 \\
Thermal Relaxation & 13291.67 & 14245.42 & 12446.39 & 13505.17 & 13630.71 \\ \hline
\end{tabular}
\end{table}

\section{Conclusion}
In this paper, a novel QACO algorithm with the K-means clustering method has been proposed to overcome the disadvantages of the current quantum computing hardware, lacking of the number of the qubits and the noise in the qubit, applying on the discrete
optimization problem, like TSP. Two typical data sets have been utilized for testing the performances of the algorithm. The testing results show that the QACO with the K-means clustering method is an effective optimization algorithm and exceeds the performances of
ACO. The proposed QACO also has a feature that could mitigate the effect of the noise existing in the current quantum computing hardware, greatly expanding the application area of the QACO algorithm.

\section{Acknowledgements}
This work is supported by YiWei Quantum Technology Co., Ltd.


\section{References}
\begin{thebibliography}{99}
\bibitem{Dorigo} M. Dorigo, A. Colorni and V. Maniezzo. Distributed optimization by ant colonies. {\it Proceedings of ECAL91- European Conf. on Artificial Life}, 1991.
\bibitem{Birattari1} M.Birattari, M. Dorigo and T.Stutzle. Ant colony optimization. {\it IEEE Computational Intelligence Magazine}, 2006.
\bibitem{Marchese} M Marchese, J Yang, X Shi and Y Liang. An ant colony optimization method for generalized tsp problem. {\it Progress in Natural Science Vol. 18 (11), pp. 1417-1422}, 2008.
\bibitem{Elloumi} Olfa Bali, Walid Elloumi, Ajith Abraham and Adel M. Alimi. ACO-PSO Optimization for Solving TSP Problem with GPU Acceleration, {\it International Conference on Intelligent Systems Design and Applications}, 2017.
\bibitem{Chitty} Darren M. Chitty. Applying ACO to Large Scale TSP Instances, {\it UK Workshop on Computational Intelligence}, 2017.
\bibitem{Demiral} M. F. Demiral. Ant colony optimization for a variety of classic assignment problems. {\it International Turkish World Engineering and Science Congress, Antalya}, 2017.
\bibitem{Amirhosein} Hamed Qahri Saremi and Amirhosein Meimand Kermani. Website structure improvement: Quadratic assignment problem approach and ant colony meta-heuristic technique, {\it Applied Mathematics and Computation}, Vol. 195 (1), pp. 285-298, 2008.
\bibitem{Alfredo} Alfredo Reyes Montero. Ant Colony Optimization for Solving the Quadratic Assign Problem, {\it 2015 Fourteenth Mexican International Conference on Artificial Intelligence}, 2015.
\bibitem{Doerner} K. Doerner, M. Reimann and R. F. Hartl. D-ants: savings based ants divide and conquer the vehicle routing problems. {\it Comput. Oper. Res.}, 2004.
\bibitem{Yanfang} Yanfang Ma, Jie Han, Kai Kang and Fang Yan. An Improved ACO for the Multi-depot Vehicle Routing Problem with Time Windows, {\it Proceedings of the Tenth International Conference on Management Science and Engineering Management}, 2016.
\bibitem{Patrick} John E. Bell and Patrick R. McMullen. Ant colony optimization techniques for the vehicle routing problem, {\it Advanced Engineering Informatics}, Vol. 18 (1), pp. 41-48, 2004.
\bibitem{Rizzoli} A. E. Rizzoli, R. Montemanni, E. Lucibello and L. M. Gambardella. Ant colony optimization for real-world vehicle routing problems, {\it Swarm Intelligence}, vol. 1, pp. 135-151, 2007.
\bibitem{Benioff} Paul Benioff. Quantum mechanical models of turing machines that dissipate no energy. {\it Phys. Rev. Lett.}, vol. 48(23), pp. 1581-1585, 1982.
\bibitem{Deutsch} David Deutsch. Quantum theory, the Church-Turing principle and the universal quantum computer, {\it Proceedings of the Royal Society of London Series A}, vol. 400 (1818), pp. 97-117, 1985.
\bibitem{Shor} Peter W. Shor. Algorithms for quantum computation: discrete logarithms and factoring, {\it Proceedings 35th Annual Symposium on Foundations of Computer Science}, 1994.
\bibitem{Schmidhuber} Jurgen Schmidhuber. Deep learning in neural networks: An overview, {\it Neural Networks}, vol. 61, pp. 85-117, 2015.
\bibitem{Prieto} Alberto Prieto, Beatriz Prieto and Eva Martinez Ortigosa. Neural networks: An overview of early research, current frameworks and new challenges, {\it Neurocomputing}, vol. 214, pp. 242-268, 2016.
\bibitem{McCall} John McCall. Genetic algorithms for modelling and optimisation, {\it Journal of Computational and Applied Mathematics}, vol. 184 (1), pp. 205-222, 2005.
\bibitem{Khaled} Khaled Salah Mohamed. Bio-Inspired Machine Learning Algorithm: Genetic Algorithm, {\it Machine Learning for Model Order Reduction}, pp 19-34, 2018.
\bibitem{Birattari} Marco Dorigo, Mauro Birattari and Thomas Stutzle. Ant colony optimization, {\it IEEE Computational Intelligence Magazine}, vol. 1 (4), pp. 28-39, 2006.
\bibitem{QNiu} Q. Niu, L. Wang and M Fei. A novel quantum ant colony optimization algorithm. {\it Lecture Notes in Computer Science book series (LNCS), Bio-Inspired Computational Intelligence and Applications}, vol. 4688, 2007.
\bibitem{LiP} Li P and Wang H. Quantum ant colony optimization algorithm based on bloch spherical search. {\it Neur Netw World}, vol. 22, pp. 325-341, 2012.
\bibitem{Echanobe} M. Garcia de Andoin and J. Echanobe. Implementable hybrid quantum ant colony optimization algorithm, {\it Quantum Machine Intelligence}, vol. 4 (12), 2022.
\bibitem{Neumann} F. Neumann and C. Witt. Runtime analysis of a simple ant colony optimization algorithm. {\it Algorithmica}, vol. 54 (2), 2009.
\bibitem{Panchi} Kaoping Song, Panchi Li and Erlong Yang. Quantum ant colony optimization with application. {\it Sixth International Conference on Natural Computation}, 2010.
\bibitem{MeijiaoL} Meijiao Liu, Yanhui Li and Mingyi Xia. A Two-Way Parallel Slime Mold Algorithm by Flow and Distance for the Travelling Salesman Problem, {\it Applied Sciences}, vol. 10 (18), 2020.
\bibitem{IBMQ} IBM Quantum, https://quantum-computing.ibm.com/.
\bibitem{HeLH} He-Liang Huang, Xiao-Yue Xu, etc. Near-term quantum computing techniques: Variational quantum algorithms ,error mitigation, circuit compilation, benchmarking and classical simulation. {\it Science China Physcis, Mechanics \& Astronomy}, vol. 66, 2023.
\end{thebibliography}
\end{document}